# Giant interfacial perpendicular magnetic anisotropy in MgO/CoFe/capping layer structures

Shouzhong Peng[1,2,3], Weisheng Zhao[1,2,a)], Junfeng Qiao[1,2], Li Su[1,2], Jiaqi Zhou[1,2], Hongxin Yang[4,5], Qianfan Zhang[1,6], Youguang Zhang[1,2], Cecile Grezes[3], Pedram Khalili Amiri[3,7] and Kang L. Wang[3]

[1]Fert Beijing Institute, BDBC, Beihang University, Beijing 100191, China

[2]School of Electronic and Information Engineering, Beihang University, Beijing 100191, China

[3]Department of Electrical Engineering, University of California, Los Angeles, California 90095, USA

[4]Unité Mixte de Physique, CNRS, Thales, Univ. Paris-Sud, Université Paris-Saclay, Palaiseau 91767, France

[5]Univ. Grenoble Alpes, INAC-SPINTEC, F-38000 Grenoble, France; CNRS, SPINTEC, F-38000 Grenoble, France; and CEA, INAC-SPINTEC, F-38000 Grenoble, France

[6]School of Materials Science and Engineering, Beihang University, Beijing 100191, China

[7]Inston Inc., Los Angeles, California 90095, USA

**ABSTRACT**

Magnetic tunnel junction (MTJ) based on CoFeB/MgO/CoFeB structures is of great interest due to its application in the spin-transfer-torque magnetic random access memory (STT-MRAM). Large interfacial perpendicular magnetic anisotropy (PMA) is required to achieve high thermal stability. Here we use first-principles calculations to investigate the magnetic anisotropy energy (MAE) of MgO/CoFe/capping layer structures, where the capping materials include 5$d$ metals Hf, Ta, Re, Os, Ir, Pt, Au and 6$p$ metals Tl, Pb, Bi. We demonstrate that it is feasible to enhance PMA by using proper capping materials. Relatively large PMA is found in the structures with capping materials of Hf, Ta, Os, Ir and Pb. More importantly, the MgO/CoFe/Bi structure gives rise to giant PMA (6.09 mJ/m$^2$), which is about three times larger than that of the MgO/CoFe/Ta structure. The origin of the MAE is elucidated by examining the contributions to MAE from each atomic layer and orbital. These findings provide a comprehensive understanding of the PMA and point towards the possibility to achieve advanced-node STT-MRAM with high thermal stability.

## I. INTRODUCTION

Spin-transfer-torque magnetic random access memory (STT-MRAM) is becoming a mainstream non-volatile memory with high density, infinite endurance and low power consumption.[1–7] The core device of the STT-MRAM is the magnetic tunnel junction (MTJ), which consists of a free layer and a reference layer separated by a barrier layer. A strong tunnel magnetoresistance (TMR) effect can be realized by using MgO as the barrier layer.[8,9] However, one of the main challenges for the MgO-based MTJs is how to obtain high thermal stability so that the data can be stored for a long time.[10,11] The thermal stability factor can be expressed as $\Delta = H_K M_S V / 2 k_B T$, where $H_K$ is the anisotropy field, $M_S$ is the saturation magnetization, $V$ is the volume of the free layer, $k_B$ is the Boltzmann constant and $T$ is the temperature. In order to achieve a data retention time of 10 years (the industry standard) for large array memory, the thermal stability factor required must be larger than 60.[12,13] In addition, as the memory density increases, the volume of the free layer shrinks, which will lead to a decrease of the thermal stability. As a consequence, it is essential to have large magnetic anisotropy for high-capacity and high-density STT-MRAMs.

In 2010, Ikeda *et al.* reported large interfacial perpendicular magnetic anisotropy (PMA) in the Ta/CoFeB/MgO/CoFeB/Ta structure.[2] With a lateral dimension of 40 nm, this structure showed a thermal stability factor of 43, a TMR ratio of 124% and a switching current of 49 μA. This work inspired intensive research on the PMA-based STT-MRAM, and a lot of efforts have been devoted to the PMA enhancement in the MgO/CoFeB-based structures.[14–20] In 2011, Worledge *et al.* investigated the Ta/CoFeB/MgO and Ru/CoFeB/MgO structures and showed that the Ta/CoFeB interface plays a key role to the PMA.[21] Later, several works reported large PMA in the CoFeB/MgO- or MgO/CoFeB-based structures by using Hf, Mo, Ir or Nb as the seed or capping materials.[22–27] Peng *et al.* attributed the PMA variations to the different interfacial anisotropy at the ferromagnetic layer/capping layer interface and pointed out that it is possible to tune PMA by choosing different capping materials.[28] Another method to increase the PMA is to use an MgO/CoFeB/Ta/CoFeB/MgO recording structure.[29] However, the thermal stability remains insufficient for the 20 nm diameter MTJs.[30] Therefore, it is important to explore new materials to obtain large enough PMA.

In this letter, we systematically investigate the magnetic anisotropy energy (MAE) of different MgO/CoFe/X structures, with the capping layer (X) including 5*d* metals Hf, Ta, Re, Os, Ir, Pt, Au and 6*p* metals Tl, Pb, Bi. Relatively large PMA values can be obtained with X=Hf, Ta, Os, Ir and Pb. Moreover, we report for the first time that the Mg/CoFe/Bi structure gives rise to giant PMA, promising for the development of high-density STT-MRAM.

## II. METHODOLOGY

Our first-principles calculations are performed using the Vienna *ab initio* simulation package (VASP)[31–33] with a generalized gradient approximation (GGA)[34] and the projector-augmented wave (PAW)[35,36] potentials. Figure 1 shows the crystalline structures of MgO/CoFe/X, CoFe/X (X=Hf, Ta, Re, Os, Ir, Pt, Au) and CoFe/X (X=Tl, Pb, Bi). A 15 Å vacuum layer is included on top of all the structures. Our previous work proved that the interfacial magnetic anisotropy in the MgO/CoFe/X structure came from both the MgO/CoFe and CoFe/X interfaces, and these two interfaces can be analyzed separately when the CoFe layer is thick enough (e.g. 9 monolayers).[28] Hence we focus on the MAEs of the CoFe/X structures here. The cut-off energy used in the calculations is 520 eV. The K-point meshes for CoFe/X (X=Hf, Ta, Re, Os, Ir, Pt, Au) structures and CoFe/X (X=Tl, Pb, Bi) structures are 20×20×1 and 14×14×1, respectively, which are sufficient to ensure a good convergence of the MAEs. The structural relaxations are performed until the forces on the atoms are less than 0.01 eV/Å. Previously, several works demonstrated that the bulk anisotropy in the MgO/CoFeB-based structures could be negligible.[2,22,24] Therefore, in our calculations, the in-plane lattice constants of the CoFe/X (X=Hf, Ta, Re, Os, Ir, Pt, Au) and the CoFe/X (X= Tl, Pb, Bi) structures are constrained to $a$=2.83 Å (the lattice constant of the bulk CoFe) and $\sqrt{2}a$, respectively, to eliminate the bulk contributions. Finally, the MAE is determined as the difference of the total energies when the magnetization orients along the in-plane ([100]) and perpendicular ([001]) directions, taking the spin-orbit coupling (SOC) into account.

## III. RESULTS AND DISCUSSION

Table I shows the calculated MAE results of the MgO/CoFe/X structures. We can see that different capping materials lead to very different MAEs. Relatively large PMA can be achieved by using Hf, Ta, Os, Ir and Pb as capping materials, agreeing with experimental results.[21,22,25] More interestingly, the MgO/CoFe/Bi structure shows an extremely large PMA value up to 6.09 mJ/m$^2$, which is about three times larger than that of the MgO/CoFe/Ta structure. This giant PMA makes Bi a very promising capping material for the high-density STT-MRAM.

In the following, we analyze the contributions to MAE from each atomic layer to gain insight into the origin of the magnetic anisotropy. The layer-resolved MAE method can be found in previous works.[28,37–39] Figure 2 shows the layer-resolved MAEs of CoFe/X structures with X=Ta, Re and Ir, whose 5$d$ subshells are less than half occupied, half occupied and more than half occupied, respectively. The result of the CoFe/Bi is also shown in this figure in consideration of the giant PMA in this structure. It can be seen that MAEs mainly originate from the surface and interface, which agrees with previous works.[37] The interfacial capping atoms are magnetically polarized by the CoFe layer and contribute to large MAEs due to the strong SOC, similar to the cases in CoPt and FePt ordered alloys.[40] The positive MAEs from the interfacial Ta and Ir atoms lead to PMA at CoFe/Ta and CoFe/Ir interfaces, while the negative MAE from the interfacial Re atom results in the in-plane anisotropy at the CoFe/Re interface. More importantly, the interfacial Bi atom contributes to a very large positive MAE, which explains the giant PMA in the MgO/CoFe/Bi structure.

Next we explore the origin of the interfacial magnetic anisotropy by analyzing the orbital-resolved MAEs. Within the perturbation theory, the MAE can be expressed as:[41]

$$MAE = \xi^2 \sum_{o^\sigma, u^{\sigma'}} \frac{\left|\langle o^\sigma | L_z | u^{\sigma'} \rangle\right|^2 - \left|\langle o^\sigma | L_x | u^{\sigma'} \rangle\right|^2}{\epsilon_{u^{\sigma'}} - \epsilon_{o^\sigma}} \quad (1)$$

where $\xi$ is the SOC constant, $o^\sigma$ ($u^{\sigma'}$) and $\epsilon_{o^\sigma}$ ($\epsilon_{u^{\sigma'}}$) denote eigenstates and eigenvalues, respectively, of the occupied (unoccupied) states with spin $\sigma$ ($\sigma'$). The $L_z$ ($L_x$) represents the angular-momentum operator with magnetization along the $z$ ($x$) direction. Hence, one can obtain the contributions to MAE from the hybridizations between different orbitals.[28,39] For the interfacial Ta atom, the contributions to MAE from the $p$ orbitals are larger than those from the $d$ orbitals, as shown in Fig. 3(a) and 3(b). Though the hybridization between $p_y$ and $p_x$ orbitals contributes to in-plane anisotropy, the positive contribution from the hybridization between $p_y$ and $p_z$ orbitals leads to large PMA for the CoFe/Ta structure. Figure 3(c) and 3(d) exhibit the contributions to MAE from the *p* and *d* orbitals of the interfacial Re atom, respectively. The hybridizations between $p_y$ and $p_z$, $d_{yz}$ and $d_{z^2}$, $d_{xy}$ and $d_{xz}$ orbitals contribute to relatively large negative MAEs, resulting in the in-plane anisotropy. For the interfacial Ir atom, the contributions from *p* orbitals are negligible, while large contributions can be found from the *d* orbitals, as shown in Fig. 3(e) and 3(f). In addition, the positive contribution from the hybridization between $d_{yz}$ and $d_{z^2}$ orbitals leads to the large PMA in the CoFe/Ir structure.

Considering the giant PMA in the CoFe/Bi structure, we focus on the analysis of the MAE at the CoFe/Bi interface in the following. The orbital-resolved MAEs of the interfacial Bi atom in the CoFe/Bi system are shown in Fig. 4(a) and 4(b). The contributions from the *p* orbitals are much larger than those from the *d* orbitals. Moreover, an extremely large positive MAE from the hybridization between $p_y$ and $p_z$ orbitals can be found in Fig. 4(a), which explains the giant PMA from the interfacial Bi atom. In order to further study the origin of this PMA, we calculate the projected density of states (PDOS) and the charge density difference. The PDOS of the surface (interface) Bi atom in the Bi slab and the CoFe/Bi structure are shown in Fig. 4(c) and 4(d), respectively. It is clear that the presence of the CoFe layer strongly affects the PDOS of the Bi atom. Great changes of the $p_z$ peaks at the Fermi energy can be found, which have a large influence on the MAE. Figure 4(e) and 4(f) illustrate the top view and side view, respectively, of the charge density difference at the CoFe/Bi interface. We can see obvious redistribution of the Bi-$p_z$ orbital and the Co-$d_{z^2}$, Co-$d_{xz}(d_{yz})$ orbitals, which reveals the strong hybridization between interfacial Bi-*p* orbitals and Co-*d* orbitals. This strong hybridization together with the strong SOC of Bi atom lead to the giant PMA in the CoFe/Bi structure.

## IV. CONCLUSION

In conclusion, first-principles calculations were performed to study the interfacial MAEs of the MgO/CoFe/X structures, where the capping layer (X) includes 5*d* metals Hf, Ta, Re, Os, Ir, Pt, Au and 6*p* metals Tl, Pb, Bi. The results showed that it is feasible to enhance PMA by using proper capping materials. Relatively large PMA was observed with X=Hf, Ta, Os, Ir and Pb, agreeing with experimental results. More importantly, the MgO/CoFe/Bi structure gives rise to the giant PMA up to 6.09 mJ/m$^2$, which is about three times larger than that of the MgO/CoFe/Ta structure. The origin of PMA was investigated by analyzing the contributions to MAE from each atomic layer and orbital. We found that the MAEs mainly come from the interfaces and surfaces, consistent with previous reports. Moreover, the strong hybridization between Co and Bi together with the strong SOC of Bi lead to the giant PMA at the CoFe/Bi interface, making it possible to achieve high thermal stability for STT-MRAMs with high capacity and high density.

**ACKNOWLEDGMENTS**

The authors thank Albert Fert and Mairbek Chshiev for fruitful discussions. S.Z.P. thanks the support by the China Scholarship Council (CSC). W.S.Z. thanks the support by the International Collaboration 111 Project B16001 from the Ministries of Education and Foreign Experts, the International Collaboration Project (2015DFE12880) from the Ministry of Science and Technology in China, the National Natural Science Foundation of China (Grant No. 61471015 and No. 61571023) and the Beijing Municipal of Science and Technology (Grant No. D15110300320000). H.X.Y. acknowledges the support by the ANR projects ULTRASKY and SOSPIN. K.L.W. acknowledges the support by the National Science Foundation under award #ECCS1611570 and the support by the Spins and Heat in Nanoscale Electronic Systems (SHINES), an Energy Frontier Research Center funded by the US Department of Energy (DOE), Office of Science, Basic Energy Sciences (BES) under award #DE-SC0012670. The work was partly supported by Inston Inc. through a Phase II Small Business Innovation Research award from the National Science Foundation.

**FIGURES AND FIGURE LEGENDS**

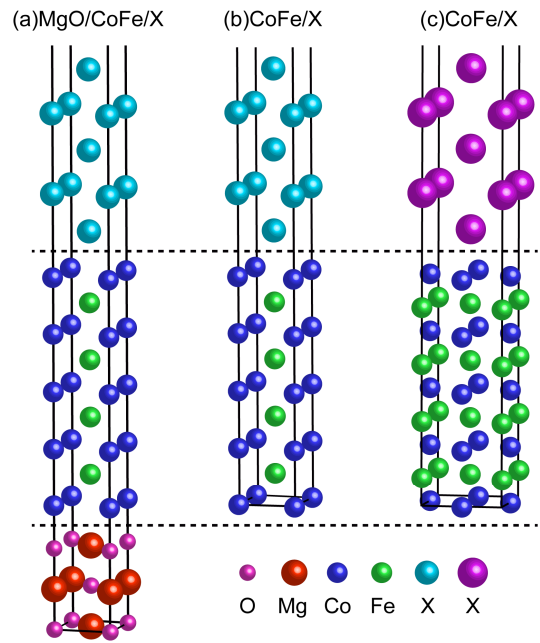

FIG. 1. Schematics of crystalline structures for (a) MgO/CoFe/X, (b) CoFe/X (X=Hf, Ta, Re, Os, Ir, Pt, Au) and (c) CoFe/X (X=Tl, Pb, Bi). A 15 Å vacuum layer is included on top of all these structures.

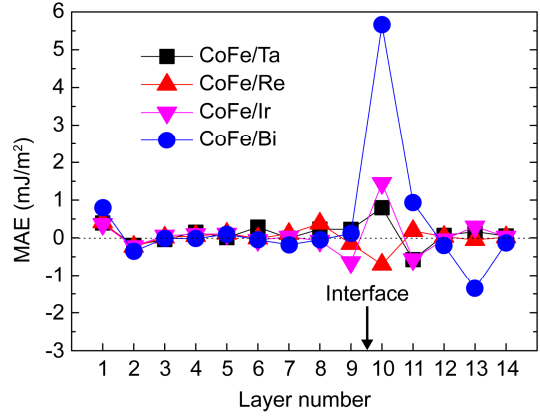

FIG. 2. Layer-resolved magnetic anisotropy energies (MAEs) of different CoFe/X (X=Ta, Re, Ir and Bi) structures. Nine CoFe monolayers, five X monolayers and a 15 Å vacuum layer are included in the structures (as shown in Fig. 1(b) and 1(c)). MAEs mainly originate from the surface and interface. Moreover, the large positive MAE from the interfacial Bi atom leads to the giant perpendicular magnetic anisotropy (PMA) in the Bi-capped structure.

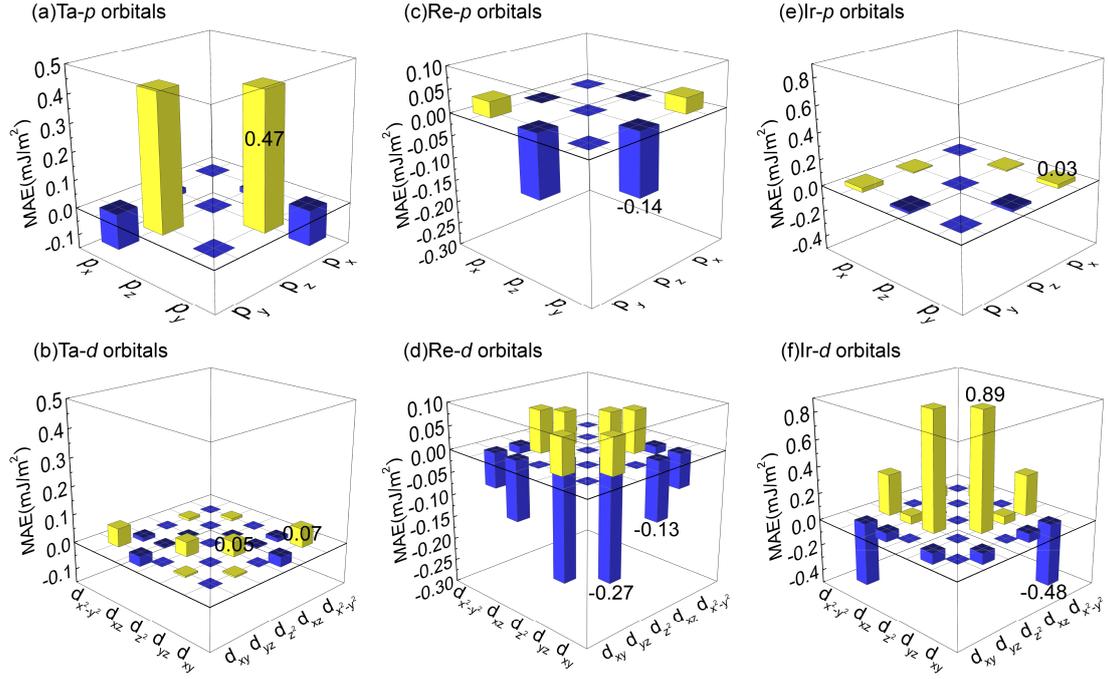

FIG. 3. Orbital-resolved MAEs of interfacial X (X=Ta, Re, Ir) atoms in the CoFe/X systems. Contributions to MAE from (a) Ta-$p$ orbitals, (b) Ta-$d$ orbitals, (c) Re-$p$ orbitals, (d) Re-$d$ orbitals, (e) Ir-$p$ orbitals and (f) Ir-$d$ orbitals. The yellow (blue) bars represent the positive (negative) MAEs. For the interfacial Ta and Ir atoms, the PMA mainly originates from the hybridization between $p_y$ and $p_z$ orbitals, and hybridization between $d_{yz}$ and $d_{z^2}$ orbitals, respectively. For the interfacial Re atom, the in-plane anisotropy mainly originates from hybridizations between $p_y$ and $p_z$, $d_{yz}$ and $d_{z^2}$, $d_{xy}$ and $d_{xz}$ orbitals.

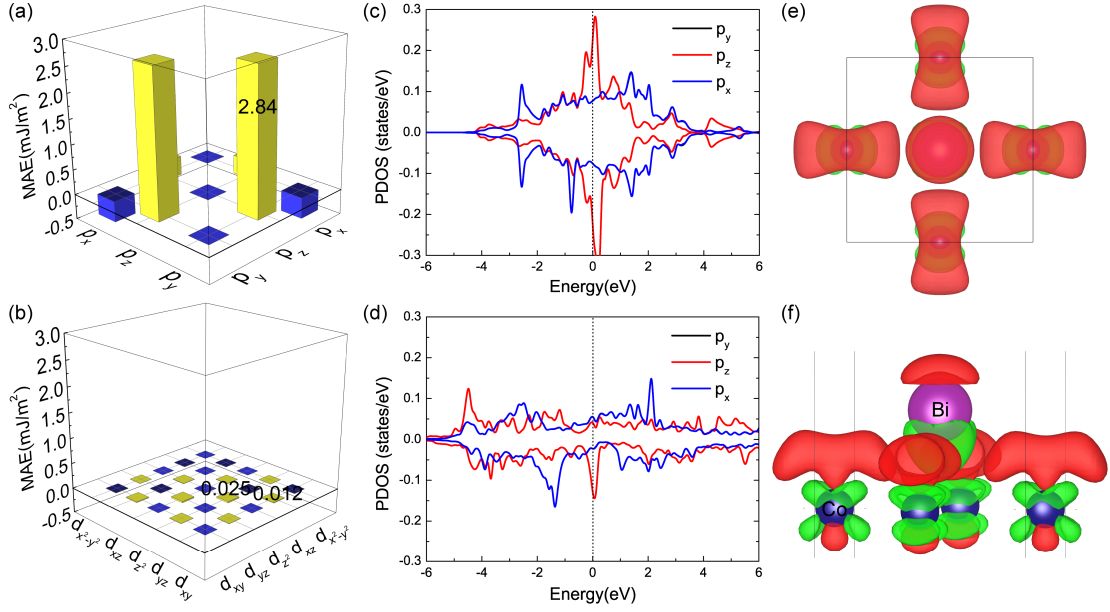

FIG. 4. (a), (b) Orbital-resolved MAEs of the interfacial Bi atom in the CoFe/Bi system. Contributions to MAE from (a) Bi-$p$ orbitals and (b) Bi-$d$ orbitals. The large positive MAE from the hybridization between $p_y$ and $p_z$ orbitals explains the giant PMA from the interfacial Bi atom. (c), (d) Projected density of states (PDOS) of the surface (interface) Bi atom in the Bi slab and the CoFe/Bi structure, respectively. The $p_x$ and $p_y$ orbitals are degenerate by the structural symmetry. The vertical dashed lines mark the Fermi level. Great changes can be found for the Bi-$p$ orbitals due to the presence of the CoFe layer. (e) Top view and (f) side view of the charge density difference ($\Delta\rho = \rho(CoFe/Bi) - \rho(CoFe) - \rho(Bi)$) at the CoFe/Bi interface with the isosurface level $\Delta\rho = \pm 0.03 e/Å^3$. The red (green) clouds denote the charge accumulation (depletion), while the blue (magenta) balls denote the Co (Bi) atoms. The redistribution of the Bi-$p_z$ orbital and the Co-$d_{z^2}$, Co-$d_{xz}(d_{yz})$ orbitals reveals the strong hybridization between interfacial Bi-$p$ orbitals and Co-$d$ orbitals.

TABLE I. Calculated MAE results of different MgO/CoFe/X structures. Relatively large PMA can be achieved by using Hf, Ta, Os, Ir and Pb as capping materials. Moreover, the MgO/CoFe/Bi structure gives rise to a giant PMA value.

| X | Hf | Ta | Re | Os | Ir | Pt | Au | Tl | Pb | Bi |
|---|---|---|---|---|---|---|---|---|---|---|
| MAEs of MgO/CoFe/X structures (mJ/m$^2$) | 2.22 | 1.70 | 0.32 | 1.42 | 0.98 | 0.74 | 0.17 | -2.02 | 1.26 | 6.09 |